\documentclass[preprint,superscriptaddress,aps]{revtex4}
\usepackage{amsfonts}
\usepackage{graphics}
\usepackage{graphicx}
\usepackage{slashed}
\newcommand{\be}{\begin{equation}}
\newcommand{\ee}{\end{equation}}
\newcommand{\en}{\end{equation}}
\newcommand{\ba}{\begin{eqnarray}}
\newcommand{\ea}{\end{eqnarray}}
\newcommand{\bea}{\begin{eqnarray}}
\newcommand{\eea}{\end{eqnarray}}
\newcommand{\bq}{\begin{eqnarray}}
\newcommand{\eq}{\end{eqnarray}}

\def\pls{\partial\!\!\!/}

\def\ps{p\!\!\!/}

\def\bs{\slashed{b}}

\def\sech{\mathrm{sech}}
\def\sinh{\mathrm{sinh}}
\def\tanh{\mathrm{tanh}}

\begin{document}

\title{Ambiguities of the CPT-even aether-like Lorentz-breaking term at the finite temperature}
\author{T. Mariz}
\affiliation{Instituto de F\'\i sica, Universidade Federal de Alagoas, 57072-270, Macei\'o, Alagoas, Brazil}
\email{wserafim,tmariz@fis.ufal.br}
\author{J. R. Nascimento}
\affiliation{Departamento de F\'{\i}sica, Universidade Federal da Para\'{\i}ba\\
 Caixa Postal 5008, 58051-970, Jo\~ao Pessoa, Para\'{\i}ba, Brazil}
\email{jroberto, petrov@fisica.ufpb.br}
\author{A. Yu. Petrov}
\affiliation{Departamento de F\'{\i}sica, Universidade Federal da Para\'{\i}ba\\
 Caixa Postal 5008, 58051-970, Jo\~ao Pessoa, Para\'{\i}ba, Brazil}
\email{jroberto, petrov@fisica.ufpb.br}
\author{W. Serafim}
\affiliation{Instituto de F\'\i sica, Universidade Federal de Alagoas, 57072-270, Macei\'o, Alagoas, Brazil}
\email{wserafim,tmariz@fis.ufal.br}
\begin{abstract}
In this paper, we consider the finite temperature behaviour of the CPT-even aether-like Lorentz-breaking term in the extended Lorentz-breaking QED and demonstrate that its ambiguities whose presence has been shown earlier in the zero temperature case stay also at the finite temperature.
\end{abstract}

\maketitle

\section{Introduction}
\label{intro}

The problem of ambiguities is known to be the central one for the Lorentz-breaking theories. Already in 1999, the seminal paper by Jackiw \cite{JackAmb} established the deep reasons for the ambiguous results for the perturbative corrections in the Lorentz-breaking QED. The ambiguity of the Carroll-Field-Jackiw term \cite{CFJ}, discussed in details in \cite{Chung}, became a paradigmatic example. Further, the ambiguity of results was shown to occur not only in the usual, zero temperature case, but also at the finite temperature (see f.e. \cite{Seminara}). The detailed discussion of ambiguities of the Carroll-Field-Jackiw term at the finite temperature has been presented in \cite{ours,ours1}.

At the same time, the Carroll-Field-Jackiw term is not an unique term displaying ambiguity of results at the quantum level. The dependence of results on the regularization scheme has been showed, first, for the aether-like term \cite{aether}, second, for the higher-derivative Lorentz-breaking term \cite{MNP} representing itself as a linear combination of the higher-derivative CFJ-like term and the Myers-Pospelov term \cite{MP}. Recently, the perturbative generation of this term in the finite temperature case has been also discussed \cite{LMS}. The issues related to the unitarity in the theory involving such a term have been considered in \cite{Reyes}.

However, the main object of interest for us is the CPT-even Lorentz-breaking term, that is, the aether term, proposed in \cite{Kostel} as a possible ingredient of the extended standard model, discussed in \cite{Carroll} within the higher dimensions context, and applied in \cite{Almeida} within the brane context. Within our studies, we will be interested mostly in the Lorentz-breaking QED with an additive aether term which is known to be compatible with the gauge symmetry. Many issues related with the presence of this term at the classical level, especially examples of exact solutions and dispersion relations in corresponding field theories, have been discussed in \cite{aethercl}. Also, it worth to notice that this term emerges within the dual embedding procedure \cite{Clovis}. However, its quantum aspects are studied less that those one for the CFJ term. The main result found for the aether term at the quantum level is that its perturbative generation in a model most used for this purpose, that is, the Lorentz-breaking QED with an extra magnetic coupling, involves two different ambiguities \cite{aether1}. So, the natural question is the behaviour of both these ambiguities in the finite temperature case, comparing thus the situation for the aether term with the situation for the CFJ term \cite{Seminara,ours,ours1}. This is the problem we study in this paper.

\section{Generation of the aether term}
\label{sec:1}

We start with the extended spinor QED whose action involves both minimal and nonminimal couplings (proportional to $e$ and $g$ respectively)  and an axial term in the fermionic sector \cite{aether1}:
\begin{equation}
\label{mcn}
{\cal L}=\bar{\psi}\left[i \pls- \gamma^{\mu}(eA_{\mu}+g\epsilon_{\mu\nu\lambda\rho}F^{\nu\lambda}b^{\rho}) - m -  \gamma_{5}\bs\right]\psi.
\end{equation}
Many aspects related to this model at the zero temperature have been discussed in \cite{aether1,MNP}. In particular, it was mentioned there that namely this model allows of generation of finite one-loop contributions to the aether term and higher-derivative Lorentz-breaking terms. Therefore, let us discuss this model, or, to be more precise, the aether-like contributions, at the finite temperature.

The one-loop effective action for this model is given by
\begin{eqnarray}
S_{eff}[b,A]&=&-i\,{\rm Tr}\,\ln(i\partial\!\!\! /-e\gamma^{\mu}A_{\mu}-g\epsilon_{\mu\nu\lambda\rho}\gamma^{\mu}F^{\nu\lambda}b^{\rho}- 
m - \gamma_5 \bs).
\end{eqnarray}
In the paper, we will obtain the lower CPT-even contributions to this effective action.

\subsection{Nonminimal contribution}

The correction of the second order in the Lorentz-breaking vector $b_{\mu}$ in a purely nonminimal sector, where $e=0$, has been discussed in details in \cite{aether,aether1}. It was shown there that this correction looks like
\begin{eqnarray}
\label{s2d40}
S_{FF}(p)&=&-\frac{g^2}{2}\epsilon^{\alpha\beta\gamma\delta}\epsilon^{\alpha'\beta'\gamma'\delta'}
b_{\alpha}F_{\beta\gamma}(p)
b_{\alpha'}
F_{\beta'\gamma'}(-p) I_{\delta\delta'}
\end{eqnarray}
where
\begin{eqnarray}\label{Idelta}
I_{\delta\delta'}&=&
\int\frac{d^4k}{(2\pi)^4}\frac{1}{(k^2-m^2)^2}{\rm tr}
\big[m^2\gamma_\delta\gamma_{\delta'}+k^\mu k^\nu\gamma_\mu\gamma_\delta\gamma_\nu 
\gamma_{\delta'}\big], \nonumber\\
&=&4\int\frac{d^4k}{(2\pi)^4}\frac{1}{(k^2-m^2)^2} (g_{\delta\delta'}m^2+2k_\delta k_{\delta'}-g_{\delta\delta'}k^2).
\end{eqnarray}

At the zero temperature, this contribution yields the result
\begin{eqnarray}
\label{s2d4}
\label{ff}
S_{FF}(p)&=&C_0\,g^2m^2(b^\alpha F_{\alpha\beta})^2,
\end{eqnarray}
where the constant $C_0$ is known to be equal either to $\frac{1}{4\pi^2}$ or to zero, see \cite{aether,aether1}. This is just the aether term proposed in \cite{Carroll}. 

In order to implement the finite temperature, we take the Eq.~(\ref{Idelta}) and change it from Minkowski space to Euclidean space. For this, we must perform the following procedure: $k_0 \rightarrow i k_0$ ($g^{\mu\nu}\rightarrow -\delta^{\mu\nu}$), $d^4k\rightarrow i d^4k_E$, and $k^2\rightarrow -k_0^2-\vec k^2 = -k^2_E$, so that, we obtain
\begin{eqnarray}
\label{idelta}
I^{\delta\delta'}&=&
4i\int\frac{d^4k_E}{(2\pi)^4}\frac{1}{(k_E^2+m^2)^2}(-\delta^{\delta\delta'}m^2+2k_E^\delta k_E^{\delta'}-\delta^{\delta\delta'}k_E^2).
\end{eqnarray}
Within the first manner of calculations, we separate the space and time components of the four-momentum $k_E^\delta=(k_0,\vec{k})$, as $k_E^\delta\rightarrow \hat{k}^\delta +k_0\delta^{\delta 0}$, so that $ \hat{k}^\delta=(0,\vec{k})$. This way of calculation is a reminiscence of the paper \cite{Seminara}. Also, due to the symmetry of the integral under spacial rotations, it is possible to use the substitutions
\begin{eqnarray}
\label{symm}
\hat{k}^\alpha \hat{k}^\beta \rightarrow \frac{\hat{k}^2}{d}(\delta^{\alpha\beta}-\delta^{\alpha 0} \delta^{\beta 0}),
\end{eqnarray}
where we have promoted the $3$-dimensional space to $d$ dimensions. Therefore, by introducing also an arbitrary parameter $\mu$, to keep the mass dimension unchanged, we have
\begin{eqnarray}
I^{\delta\delta'}&=&
4i\int\frac{dk_0}{2\pi}(\mu^2)^{\frac32-\frac d2}\int\frac{d^dk}{(2\pi)^d}\frac{1}{(k_E^2+m^2)^2}(-\delta^{\delta\delta'}m^2+2k_E^\delta k_E^{\delta'}-\delta^{\delta\delta'}k_E^2) \nonumber\\
&=&8i\int\frac{dk_0}{2\pi}(\mu^2)^{\frac32-\frac d2}\int\frac{d^dk}{(2\pi)^d}\frac{1}{(\vec k^2+k_0^2+m^2)^2}\left[\frac{\vec k^2}{d}(\delta^{\delta\delta'}-\delta^{\delta0}\delta^{\delta'0})+k_0^2\,\delta^{\delta0}\delta^{\delta'0}\right] \nonumber\\
&&-4i\int\frac{dk_0}{2\pi}(\mu^2)^{\frac32-\frac d2}\int\frac{d^dk}{(2\pi)^d}\frac{1}{(\vec k^2+k_0^2+m^2)}\delta^{\delta\delta'}.
\end{eqnarray}
Then, by calculating the integrals over the space components $\vec k$, we obtain
\begin{eqnarray}
\label{Idelta2}
I^{\delta\delta'}&=&
-i2^{2-d}\pi^{-d/2}(\mu^2)^{\frac32-\frac d2}\Gamma\left(1-\frac d2\right)\delta^{\delta0}\delta^{\delta'0}\nonumber\\
&&\times\int\frac{dk_0}{2\pi}(k_0^2+m^2)^{\frac d2-2}(m^2+(d-1)k_0^2).
\end{eqnarray}
If we now calculate the $k_0$ integral, $I^{\delta\delta'}=0$, which reproduces one of the zero temperature results found in \cite{aether}.

Let us now employ the Matsubara formalism, which consist in taking $k_0=(n+1/2)2\pi/\beta$ and changing $(1/2\pi)\int dk_0\to 1/\beta \sum_n$. However, we cannot readily take the limit $d\to3$ in the Eq.~(\ref{Idelta2}), because the sum exhibits singularities. Thus, in order to isolate these singularities, let us use an explicit representation for the sum over the Matsubara frequencies \cite{For}, given by
\begin{equation}
\label{sum}
\sum_n\bigl[(n+\eta)^2+\xi^2\bigl]^{-\lambda} = \frac{\sqrt{\pi}\Gamma(\lambda-1/2)}{\Gamma(\lambda)(\xi^2)^{\lambda-1/2}}+4\sin(\pi\lambda)f_\lambda(\xi,\eta)
\end{equation}
where
\begin{equation}
\label{f}
f_\lambda(\xi,\eta) = \int^{\infty}_{|\xi|}\frac{dz}{(z^2-\xi^2)^{\lambda}}Re\Biggl(\frac{1}{e^{2\pi(z+i\eta)}-1}\Biggl),
\end{equation}
which is valid for $Re\,\lambda<1$, aside from the poles at $\lambda=1/2,-1/2,\cdots$.  Actually, $\xi=\frac{m}{2\pi T}$, and $\eta=1/2$ while all propagators are fermionic. Therefore, applying these results for the expression (\ref{Idelta2}), we obtain
\begin{equation}
\label{idd}
I^{\delta\delta'} = \frac{im^2}{\pi^2\xi^2}\delta^{\delta0}\delta^{\delta'0} F_1(\xi),
\end{equation}
where
\begin{equation}\label{F1}
F_1(\xi)=\int_{|\xi|}^\infty dz \frac{(2z^2-\xi^2)}{(z^2-\xi^2)^{1/2}}(1-\tanh(\pi z)).
\end{equation}
At high temperature limit, the above expression is temperature dependent, which can be rewrite as
\begin{equation}
I^{\delta\delta'} = \frac{i}{3} T^2 \delta^{\delta 0} \delta^{\delta' 0} + {\cal}O\left(\frac{m}{T}\right).
\end{equation}

Another manner to obtain the contribution to the two-point function, that is, to find $I^{\delta\delta'}$, is based on applying the Matsubara formalism to the expression (\ref{Idelta}), or, as is the same, (\ref{idelta}), without the symmetrization (\ref{symm}).
Again, at the finite temperature case, we carry out the Wick rotation and the discretization of the zeroth coordinate by the rule $k_0\to2\pi T(n+1/2)$, with $T=1/\beta$, so that we arrive at the following form for $I_{\delta\delta'}$
\bea
\label{i0}
I_{\delta\delta'}&=&
iT\int\frac{d^3k}{(2\pi)^3}\sum\limits_{n=-\infty}^{\infty}\frac{1}{[4\pi^{2} T^{2}{(n+1/2)}^2+\vec{k}^2+m^2]^2}
\nonumber\\&&\times
{\rm tr}
\big[m^2\gamma_\delta\gamma_{\delta'}-4\pi^{2} T^{2}{(n+1/2)}^2\gamma_0\gamma_\delta\gamma_0
\gamma_{\delta'}+k^ik^j\gamma_i\gamma_\delta\gamma_j
\gamma_{\delta'}\big].
\eea
It remains to find a trace. We use the fact that the $4\times 4$ Dirac matrices yield the following relations
\bea
&&{\rm tr}(\gamma^a\gamma^b\gamma^c\gamma^d)=4(\eta^{ab}\eta^{cd}-\eta^{ac}\eta^{bd}+\eta^{ad}\eta^{bc});\nonumber\\
&&{\rm tr}(\gamma^a\gamma^b)=4\eta^{ab}.
\eea
We have three situations 

(i) $\delta,\delta'=k,l$ (both free indices are spatial). We have
\bea
\label{ikl}
I_{kl}&=&
iT\int\frac{d^3k}{(2\pi)^3}\sum\limits_{n=-\infty}^{\infty}\frac{1}{[4\pi^{2} T^{2}{(n+1/2)}^2+\vec{k}^2+m^2]^2}
\nonumber\\&&\times
{\rm tr}
\big[m^2\gamma_k\gamma_l-4\pi^{2} T^{2}{(n+1/2)}^2\gamma_0\gamma_k\gamma_0
\gamma_l+k^ik^j\gamma_i\gamma_k\gamma_j
\gamma_l\big],\nonumber\\
&=&
-4iT\delta_{kl}\int\frac{d^dk}{(2\pi)^d}\sum\limits_{n=-\infty}^{\infty}\frac{1}{[4\pi^{2} T^{2}{(n+1/2)}^2+\vec{k}^2+m^2]^2}
\nonumber\\&&\times
\left[m^2+4\pi^{2} T^{2}{(n+1/2)}^2+\left(\frac{d-2}{d}\right)\vec{k}^2\right].
\eea
Here we replaced the spatial dimension by $d$, in order to verify the presence of ambiguities.

To evaluate this expression, we use the identity
\bea
\int\frac{d^d\vec{k}}{(2\pi)^d}\frac{a\vec{k}^2+M_{n}^{2}}{(\vec{k}^2+M^2_n)^2}=
\frac{1}{(4\pi)^{d/2}}\Big[\frac{d}{2}\frac{a\Gamma(1-d/2)}{(M^2_n)^{1-d/2}}+\frac{M_{n}^{2}\Gamma(2-d/2)}{(M^2_n)^{2-d/2}}
\Big].
\eea
It follows from the expression (\ref{ikl}) that in our case we must choose $M^2_n=m^2+4\pi^2T^2(n+1/2)^2$, and $a=\frac{d-2}{d}$. In this case, one has 
\bea
I_{kl}=-\frac{4iT}{(4\pi)^{d/2}}\delta_{kl}\sum\limits_{n=-\infty}^{\infty}\frac{1}{(M^2_n)^{1-d/2}}\left(\frac{ad}{2}+1-\frac{d}{2}\right)=0,
\eea
and this zero result matches the case achieved in one of the regularizations in \cite{aether} at the zero temperature, so, at least in one of the regularizations this part of the two-point function vanishes. We note that if we choose $d=3$ from the very beginning, the result also will be zero. Effectively it means that the introduction of the finite temperature plays the role of a specific regularization different from those ones used in \cite{aether} and yielding the zero result (in particular, in the zero temperature limit).

(ii) $\delta=i,\delta'=0$, or vice versa. In this case all traces in (\ref{i0}) straightforwardly vanish, and the result is zero.

(iii) $\delta=0,\delta'=0$. 
Let us consider the expression (\ref{Idelta}) for the case $\delta=\delta'=0$. First, we impose here $\delta=\delta'=0$ and find
\begin{eqnarray}
I_{00}&=&
4\int\frac{d^4k}{(2\pi)^4}\frac{1}{(k^2-m^2)^2} [-g_{00}(k^2-m^2)+2k_0^2].
\end{eqnarray}
Then we carry out Wick rotation and take into account that $g_{00}=1$, we get
\begin{eqnarray}
I_{00}&=&
4i\int\frac{dk_{0}d^3\vec{k}}{(2\pi)^4} \left[\frac{1}{k^2_{0}+\vec{k}^2+m^2}-2\frac{k^2_{0}}{(k^2_{0}+\vec{k}^2+m^2)^2}\right].
\end{eqnarray}
Then we integrate over $d^3\vec{k}$:
\begin{eqnarray}
I_{00}&=&
\frac{4i}{(4\pi)^{3/2}}\int\frac{dk_{0}}{2\pi} \left[\frac{\Gamma(-1/2)}{(k^2_{0}+m^2)^{-1/2}}-2\frac{k^2_{0}\Gamma(1/2)}{(k^2_{0}+m^2)^{1/2}}\right].
\end{eqnarray}
Now, we change integration over $k_0$ by a discrete sum, with $\int dk_0\to 2\pi T\sum_n$, and $k^2_0\to 4\pi^2T^2(n+1/2)^2$, as usual. Thus, by introducing $M^2_n=m^2+4\pi^2T^2(n+1/2)^2$, we can write
\begin{eqnarray}
I_{00}&=&
\frac{4i}{(4\pi)^{3/2}}T\sum\limits_{n=-\infty}^{\infty} \left[\frac{\Gamma(-1/2)}{(M^2_n)^{-1/2}}-2\Gamma(1/2)\frac{(M^2_n-m^2)}{(M^2_n)^{1/2}}\right].
\end{eqnarray}
Since $\Gamma(1/2)=-1/2\Gamma(-1/2)$, we have
\begin{eqnarray}
I_{00}&=&
\frac{4i\Gamma(-1/2)}{(4\pi)^{3/2}}T\sum\limits_{n=-\infty}^{\infty} \left[\frac{1}{(M^2_n)^{-1/2}}+\frac{(M^2_n-m^2)}{(M^2_n)^{1/2}}\right]\nonumber\\
&=&
\frac{4i\Gamma(-1/2)}{(4\pi)^{3/2}}T\sum\limits_{n=-\infty}^{\infty} \left[\frac{2}{(M^2_n)^{-1/2}}-\frac{m^2}{(M^2_n)^{1/2}}\right].
\end{eqnarray}
We introduce $\xi=\frac{m}{2\pi T}$, so, $M^2_n=4\pi^2T^2[\xi^2+(n+1/2)^2]$. Thus,
\bea
I_{00}&=&
\frac{4i\Gamma(-1/2)}{(4\pi)^{3/2}}T\sum\limits_{n=-\infty}^{\infty} \left\{\frac{4\pi T}{[\xi^2+(n+1/2)^2]^{-1/2}}-\frac{m^2}{2\pi T[\xi^2+(n+1/2)^2]^{1/2}}\right\}.
\eea
To sum over $n$, we use the formula (\ref{sum}). In our case, $\lambda=1/2$ or $\lambda=-1/2$. The integral in (\ref{sum}) evidently converges in both these cases, the only dangerous terms are those one involving Euler gamma functions. To avoid the difficulty, let us temporarily introduce the small positive $\epsilon$ parameter in the expression for $I_{00}$:
\bea
I_{00}&=&-
\frac{i}{\pi}T\sum\limits_{n=-\infty}^{\infty} \left\{\frac{4\pi T}{[\xi^2+(n+1/2)^2]^{-1/2+\epsilon}}-\frac{m^2}{2\pi T[\xi^2+(n+1/2)^2]^{1/2+\epsilon}}\right\}.
\eea
Using the formula (\ref{sum}), we have $I_{00}=I_{00}^{(a)}+I_{00}^{(b)}$, with
\bea
I_{00}^{(a)}&=&-
\frac{i}{\pi}T \left[\frac{4\pi T\sqrt{\pi}\Gamma(-1+\epsilon)}{\Gamma(-1/2)(\xi^2)^{-1+\epsilon}}-
\frac{m^2\sqrt{\pi}\Gamma(\epsilon)}{2\pi T\Gamma(1/2)(\xi^2)^{\epsilon}}\right],\\
I_{00}^{(b)}&=&\frac{i}{\pi}T\left[
8\pi T\int_{|\xi|}^{\infty}\frac{dz}{(z^2-\xi^2)^{-1/2}}(1-\tanh(\pi z))\right.\nonumber\\
&&\left.-\frac{m^2}{\pi T}\int_{|\xi|}^{\infty}\frac{dz}{(z^2-\xi^2)^{1/2}}(1-\tanh(\pi z))\right].
\eea
It is clear that the singularity can emerge only from $I^{(a)}_{00}$. Let us study it in more detail. First of all, let us recall that $\xi=\frac{m}{2\pi T}$. Then, we have
\bea
I_{00}^{(a)}&=&-
\frac{i}{\pi^2}\frac{m^2\sqrt{\pi}}{(\xi^2)^{\epsilon}} \left[\frac{\Gamma(-1+\epsilon)}{\Gamma(-1/2)}-
\frac{\Gamma(\epsilon)}{2\Gamma(1/2)}\right].
\eea
Then, we took into account that $\Gamma(-1/2)=-2\Gamma(1/2)$. We have
\bea
I_{00}^{(a)}&=&-
\frac{i}{\pi^2}\frac{m^2\sqrt{\pi}}{(\xi^2)^{\epsilon}} \left[\frac{\Gamma(-1+\epsilon)+\Gamma(\epsilon)}{\Gamma(-1/2)}\right].
\eea
And $\Gamma(-1+\epsilon)+\Gamma(\epsilon)=-1$ (just the same relation between Euler  gamma functions implied cancellation of divergences in the zero temperature case  \cite{aether}). We can tend $\epsilon\to 0$ and hence put $(\xi^2)^{\epsilon}=1$. So,
\bea
\label{i00fin}
I_{00}^{(a)}&=&
\frac{i}{\pi^2}\frac{m^2\sqrt{\pi}}{\Gamma(-1/2)}=-\frac{im^2}{2\pi^2}.
\eea
Thus, the complete result for $I_{00}$ is
\bea
\label{i00}
I_{00}&=&-\frac{im^2}{2\pi^2}+i\left[
8 T^2\int_{|\xi|}^{\infty}dz(z^2-\xi^2)^{1/2}(1-\tanh(\pi z))\right.\nonumber\\
&&\left.-\frac{m^2}{\pi^2}\int_{|\xi|}^{\infty}\frac{dz}{(z^2-\xi^2)^{1/2}}(1-\tanh(\pi z))\right]\nonumber\\
&=&-\frac{im^2}{2\pi^2}+i\left[8 T^2F_2(\xi)- \frac{m^2}{\pi^2}F_3(\xi)\right],
\eea
where
\bea
\label{f2}
\label{F2}F_2(\xi)&=&\int_{|\xi|}^{\infty}dz(z^2-\xi^2)^{1/2}(1-\tanh(\pi z)),\\
\label{F3}F_3(\xi)&=&\int_{|\xi|}^{\infty}\frac{dz}{(z^2-\xi^2)^{1/2}}(1-\tanh(\pi z)).
\eea
However, in any case the terms involving the integrals in this expression are finite both at lower and upper limits. One can note that the non-vanishing zero temperature contribution (\ref{i00fin}) (and the corresponding calculation scheme) was not discussed in \cite{aether}. The main reason for it consists in the fact that this scheme requires a very especial role of the zero coordinate which can be naturally explained namely by the finite temperature prescription.

Following third manner of the calculation, we can first calculate the trace in (\ref{s2d40}) in four dimensions as it has been done in \cite{aether}, express the result in terms of $k^2$ and only afterwards implement the finite temperature. In this case we have after the Wick rotation but before introduction of the finite temperature
\bea
I_{\delta\delta'}=2i\eta_{\delta\delta'}\int\frac{d^4k_E}{(2\pi)^4}\frac{k^2_E+2m^2}{(k^2_E+m^2)^2}=2i\eta_{\delta\delta'}I,
\eea
where
\bea
I=\int \frac{d^4k_E}{(2\pi)^4}\frac{k^2_E+2m^2}{(k^2_E+m^2)^2}.
\eea
Let us calculate the integral $I$. In the calculation below, omit the index $E$ henceforth. 
First, we can write
\bea
\frac{1}{k^2+m^2}=\frac{1}{k^2+m^2}-\frac{1}{k^2}+\frac{1}{k^2}=-\frac{m^2}{k^2(k^2+m^2)}+\frac{1}{k^2}.
\eea
So, the integrand of $I$ looks like
\bea
\label{integrand}
\frac{k^2+2m^2}{(k^2+m^2)^2}=\frac{m^2}{(k^2+m^2)^2}-\frac{m^2}{k^2(k^2+m^2)}+\frac{1}{k^2}=-\frac{m^4}{k^2(k^2+m^2)^2}+
\frac{1}{k^2}.
\eea 

In the zero temperature case, with the Feynman representation, we have:
\bea
I&=&\int\frac{d^4k}{(2\pi)^4}\frac{k^2+2m^2}{(k^2+m^2)^2}=-\int\frac{d^4k}{(2\pi)^4}\frac{m^4}{k^2(k^2+m^2)^2}=
\nonumber\\&=&-2m^4\int_0^1dx\int
\frac{d^4k}{(2\pi)^4}\frac{x}{(k^2+m^2x)^3}
=-\frac{m^2}{16\pi^2},
\eea
which matches the result in \cite{aether}, since the integral from the $\frac{1}{k^2}$, the last term of (\ref{integrand}), vanishes within the dimensional regularization.
Thus, the equivalent expression for $I$ is
\bea
I=\int_0^1dx\int\frac{d^4k}{(2\pi)^4}\left[-\frac{2m^4x}{(k^2+m^2x)^3}+\frac{1}{k^2}\right].
\eea
We can present this expression as $I=I^{\prime}+I_0$, where
\bea
I^{\prime}=-2m^4\int_0^1dx\int\frac{d^4k}{(2\pi)^4}\frac{x}{(k^2+m^2x)^3},
\eea
and $I_0=-2m^4\int\frac{d^4k}{(2\pi)^4}\frac{1}{k^2}$.

Now, we implement the finite temperature just to this expression:
\bea
I^{\prime}=-2m^4T\sum\limits_{n=-\infty}^{\infty}\int_0^1dx\int\frac{d^3\vec{k}}{(2\pi)^3}\frac{x}{[\vec{k}^2+
4\pi^2T^2(n+1/2)^2+m^2x]^3}.
\eea
Integrating over spatial moment, we have
\bea
I^{\prime}=-m^4T\frac{\Gamma(3/2)}{(4\pi)^{3/2}}\sum\limits_{n=-\infty}^{\infty}\int_0^1dx\frac{x}{[4\pi^2T^2(n+1/2)^2+m^2x]^{3/2}}.
\eea
We can rewrite this expression in terms of $\xi=\frac{m}{2\pi T}$, by writing 
\bea
I^{\prime}=-\frac{m^4T}{16\pi(2\pi T)^3}\sum\limits_{n=-\infty}^{\infty}\int_0^1dx\frac{x}{[(n+1/2)^2+\xi^2x]^{3/2}}.
\eea
To factorize out the temperature dependence, we introduce new variable $\tilde{x}=\xi^2x$, so that
\bea
I^{\prime}=-\frac{m^4T}{16\pi(2\pi T)^3}\left(\frac{4\pi^2T^2}{m^2}\right)^2\sum\limits_{n=-\infty}^{\infty}\int_0^{\xi^2}d\tilde{x}\frac{\tilde{x}}{[(n+1/2)^2+\tilde{x}]^{3/2}}.
\eea
Now, in order to apply the summation formula (\ref{sum}), we must use the recurrence relation
\begin{eqnarray}\label{rc}
f_{\lambda}(\xi,\eta) &=& -\frac1{2\xi^2}\frac{2\lambda-3}{\lambda-1}f_{\lambda-1}(\xi,\eta) - \frac1{4\xi^2}\frac1{(\lambda-2)(\lambda-1)}\frac{\partial^2}{\partial \eta^2}f_{\lambda-2}(\xi,\eta),
\end{eqnarray}
because $\lambda=3/2$ is clearly out of range of validity. Therefore, this allows us to write down the result for $I^{\prime}$ as
\begin{equation}
I^{\prime} = -\frac{m^2}{16\pi^2}+\frac{T^2\pi^2}{2}F_4(\xi),
\end{equation}
where
\begin{equation}\label{F4}
F_4(\xi) = \int_0^{\xi^2}d\tilde{x}\int_{|\sqrt{\tilde{x}}|}^{\infty}dz(z^2-\tilde{x})^{1/2}\,\sech^{2}(\pi z)\tanh(\pi z).
\end{equation}

Now, let us discuss the integral $I_0$. In the zero temperature case it is certainly zero. In principle, we could treat it as zero at the finite temperature as well (actually, treating this restriction as one more regularization). However, if we implement the finite temperature, we have
\bea
I_0=T\sum\limits_{n=-\infty}^{\infty}\int\frac{d^3\vec{k}}{(2\pi)^3}\frac{1}{\vec{k}^2+
4\pi^2T^2(n+1/2)^2}.
\eea
The integral is straightforward:
\bea
I_0=-\frac{T^2}{2}\sum\limits_{n=-\infty}^{\infty}\left|n+\frac{1}{2}\right|.
\eea
Again, we change the sum over all $n$ by the sum over only non-negative $n$, so,
\bea
I_0=-T^2\sum\limits_{n=0}^{\infty}(n+1/2)=-T^2\left[\zeta(-1)+\frac{1}{2}\zeta(0)\right],
\eea
where we took into account that $\sum\limits_{n=1}^{\infty}\frac{1}{n^s}=\zeta(s)$. Since $\zeta(0)=-\frac{1}{2}$, and $\zeta(-1)=-\frac{1}{12}$, one has
\bea
I_0=\frac{T^2}{3}.
\eea
In such a case, one has 
\bea
\label{i3}
I^{\delta\delta'}=2\delta^{\delta\delta'}(I^{\prime}+I_0)=2\delta^{\delta\delta'}\left[-\frac{m^2}{16\pi^2}+\frac{T^2}{3}+
\frac{T^2\pi^2}{2}F_4(\xi)\right].
\eea

It is instructive also to discuss fourth scheme for calculation the purely nonminimal contribution to the two-point function, which has been considered in \cite{aether} here it was shown to give zero result at the zero temperature. It was shown there that in arbitrary dimension $D$ of the space-time, the expression (\ref{s2d40}), after replacement $k_{\mu}k_{\nu}\to\frac{1}{D}\eta_{\mu\nu}k^2$ and the subsequent Wick rotation, one has
\bea
I_{\delta\delta'}=2i\eta_{\delta\delta'}I,
\eea
where, unlike the previous case, now one has
\bea
I=2\int \frac{d^Dk_E}{(2\pi)^D}\frac{k^2_E(\frac{D-2}{D})+m^2}{(k^2_E+m^2)^2}.
\eea
We can implement the finite temperature into this expression. If we introduce $d=D-1$ as a purely spatial dimension (with, in principle, $d=3+\epsilon$) and then integrate over spatial components of the moment only, we get
\bea
I=\frac{2T}{(4\pi)^{d/2}(d+1)}\sum\limits_{n=-\infty}^{\infty}\left[
\frac{(d-1)\Gamma(1-\frac{d}{2})}{[m^2+4\pi^2T^2(n+\frac{1}{2})^2]^{1-\frac{d}{2}}}+
\frac{2\Gamma(2-\frac{d}{2})}{[m^2+4\pi^2T^2(n+\frac{1}{2})^2]^{2-\frac{d}{2}}}
\right].
\eea
The result of summation carried out with use of (\ref{sum}) is
\bea
I&=&\frac{(2\pi)^{d-2}T^{d-1}}{(4\pi)^{d/2}(d+1)}\left[\frac{\sqrt{\pi}}{(\xi^2)^{\frac{1}{2}-\frac{d}{2}}}
\Big((d-1)
\Gamma(\frac{1-d}{2})+
2\Gamma(\frac{3-d}{2})\Big)\right.
\nonumber\\
&&+4(d-1)\Gamma(1-\frac{d}{2})\sin (\pi(1-\frac{d}{2}))f_{1-d/2}(\xi,\frac{1}{2})\nonumber\\
&&+\left.8\Gamma(2-\frac{d}{2})
\sin (\pi(2-\frac{d}{2}))f_{2-d/2}(\xi,\frac{1}{2})
\right],
\eea
where the generic expression for the function $f_\lambda(\xi,\eta)$ is given by (\ref{f}). 

Due to the known identity $\Gamma(x+1)=x\Gamma(x)$, we can completely cancel the first (temperature independent and, actually, involving two singularities since $d=3+\epsilon$) term of this expression corroborating thus the result of \cite{aether} that within this procedure the zero temperature result exactly vanishes (we note that this mutual cancellation of singularities is also based on the Euler gamma function properties just as in \cite{aether}). We rest with
\bea
I&=&\frac{(2\pi)^{d-2}T^{d-1}}{(4\pi)^{d/2}(d+1)}\left[
4(d-1)\Gamma(1-\frac{d}{2})\sin (\pi(1-\frac{d}{2}))f_{1-d/2}(\xi,\frac{1}{2})\right.\nonumber\\
&&+8\Gamma(2-\frac{d}{2})
\sin (\pi(2-\frac{d}{2}))f_{2-d/2}(\xi,\frac{1}{2})
\Big].
\eea
Since there is no singularities more in this expression, we can put $d=3$ and, using (\ref{f2}), write down the final result as
\bea
\label{i4}
I_{\delta\delta'}=\delta_{\delta\delta'}\frac{T^2}{2}\left[2F_2(\xi)+F_3(\xi)\right].
\eea

We close this subsection giving the explicit results for the aether term obtained with use of these three procedures of calculating the $I^{\delta\delta'}$ yielding results (\ref{idd}), (\ref{i00}), (\ref{i3}) and (\ref{i4}) respectively. The corresponding expressions for the two-point function (including both the aether term and the Lorentz invariant term) are
\bea
\label{finnonmin}
S_{FF1}&=&-g^2T^2F_1(\xi)(2\vec{b}^2F_{ij}F^{ij}-4b^kF_{kj}b_iF^{ij});\nonumber\\
S_{FF2}&=&-g^2\left[-\frac{m^2}{4\pi^2}(1+4F_2(\xi))+16T^2F_3(\xi)\right](2\vec{b}^2F_{ij}F^{ij}-4b^kF_{kj}b_iF^{ij});\nonumber\\
S_{FF3}&=&g^2\left[-\frac{m^2}{16\pi^2}+\frac{T^2}{3}+\frac{T^2\pi^2}{2}F_4(\xi)\right](2b^2F_{\mu\nu}F^{\mu\nu}-4b^{\mu}F_{\mu\nu}b_{\lambda}
F^{\lambda\nu});\nonumber\\
S_{FF4}&=&g^2\frac{T^2}{2}\left[2F_2(\xi)+F_3(\xi)\right](2b^2F_{\mu\nu}F^{\mu\nu}-4b^{\mu}F_{\mu\nu}b_{\lambda}
F^{\lambda\nu}).
\eea
Here the functions $F_1(\xi)$, $F_2(\xi)$, $F_3(\xi)$, and $F_4(\xi)$ can be read off from the expressions (\ref{F1}), (\ref{F2}), (\ref{F3}), and (\ref{F4}), respectively. We close this subsection with the statement that each of the schemes implied in one of results presented in (\ref{finnonmin}) is physically consistent. However, within the first and fourth schemes the result vanishes at zero temperature, within the second one only the space-like $b^{\mu}$ contributes to the zero temperature result (which reflects the fact that introduction of the finite temperature breaks the Lorentz symmetry), and within the third one the result found in \cite{aether} is naturally promoted to the case of the presence of the finite temperature.

\subsection{Minimal contribution}

So, let us turn to the aether-like corrections essentially depending on $e$. The correction of the second order in the Lorentz-breaking vector $b_{\mu}$ in a purely minimal sector, where $g=0$, is given by the following expression:
\begin{eqnarray}
\label{1}
S_{AA}(p)&=&\frac{ie^2}{2}\int\frac{d^4k}{(2\pi)^4}\mbox{tr}(\gamma^{\mu}\frac{1}{\slashed{k}-m}\gamma^{\nu}\frac{1}{\slashed{k}+\ps-m}
\gamma_5\bs\frac{1}{\slashed{k}+\ps-m}\gamma_5\bs\frac{1}{\slashed{k}+\ps-m}+ 
\nonumber\\&&+
\gamma^{\mu}\frac{1}{\slashed{k}-m}\gamma_5\bs\frac{1}{\slashed{k}-m}
\gamma^{\nu}\frac{1}{\slashed{k}+\ps-m}\gamma_5\bs\frac{1}{\slashed{k}+\ps-m}+\nonumber\\
&&+
\gamma^{\mu}\frac{1}{\slashed{k}-m}\gamma_5\bs\frac{1}{\slashed{k}-m}
\gamma_5\bs\frac{1}{\slashed{k}-m}\gamma^{\nu}\frac{1}{\slashed{k}+\ps-m})
A_{\mu}(-p)A_{\nu}(p),
\label{ee}
\end{eqnarray}
where $p$ is an external momentum.
Let us write
\be
S(k+p)=\frac{1}{\slashed{k}+\ps-m},
\ee
which can be expanded as
\be
S(k+p)=\sum_{i=0}^{\infty} \frac{1}{\slashed{k}-m}\left( -\ps \frac{1}{\slashed{k}-m}\right)^i = \sum_{i=0}^{\infty} S_i,
\ee
where
\be
S_i\equiv \frac{1}{\slashed{k}-m}\left( -\ps \frac{1}{\slashed{k}-m}\right)^i.
\ee
The expression of eq.(\ref{ee}) can then be written as
\begin{equation}
S_{AA}(p)= \frac{ie^2}{2}A_{\mu}(-p)A_{\nu}(p)\Pi^{\mu\nu}(p) + {\cal O}(p^3),
\end{equation}
where
\bq
\Pi^{\mu\nu}(p)&=&\int\frac{d^4k}{(2\pi)^4}\mbox{tr}[\gamma^\mu S_0 \gamma_5 \bs S_0 \gamma_5 \bs S_0\gamma^\nu (S_0+S_1+S_2) \nonumber \\
&&+ \gamma^\mu S_0 \gamma_5 \bs S_0 \gamma^\nu (S_0+ S_1+S_2)\gamma_5 \bs (S_0+S_1+S_2) \nonumber \\
&&+ \gamma^\mu S_0 \gamma^\nu (S_0+S_1+S_2)\gamma_5\bs (S_0+S_1+S_2) \gamma_5\bs (S_0+S_1+S_2)].
\eq
The contributions we are interested in are quadratic in $p$. These terms are finite by power counting and, thus, ambiguity-free. 
After the expansion and the calculation of the trace, we arrive at the following expression: $\Pi^{\mu\nu}\to\Pi_1^{\mu\nu}+\Pi_2^{\mu\nu}+\Pi_3^{\mu\nu}+\Pi_4^{\mu\nu}$, with
\begin{eqnarray}
\Pi_1^{\mu\nu} &=& -\int\frac{d^4k_E}{(2\pi)^4}\frac{8}{(k_E^2+m^2)^3}[(b_E\cdot p_E)^2\delta^{\mu\nu}+(b_E\cdot p_E)p_{E}^\mu b_{E}^\nu+(b_E\cdot p_E)b_{E}^\mu p_{E}^\nu \nonumber\\
&&+p_E^2b_{E}^\mu b_{E}^\nu+b_E^2p_E^2 \delta^{\mu\nu}],
\end{eqnarray}
\begin{eqnarray}
\Pi_2^{\mu\nu} &=& \int\frac{d^4k_E}{(2\pi)^4}\frac{1}{(k_E^2+m^2)^4}[-16m^{2}(b_E\cdot p_E)^2\delta^{\mu\nu}+32k_{E}^{\mu}k_{E}^{\nu}(b_E\cdot p_E)^2+32b_{E}^{\mu}b_{E}^{\nu}(p_E\cdot k_E)^2 \nonumber \\ 
&&+24b_{E}^{2}(p_E\cdot k_E)^2\delta^{\mu\nu}+16m^{2}p_{E}^{\mu}b_{E}^{\nu}(b_E\cdot p_E)+16m^{2}b_{E}^{\mu}p_{E}^{\nu}(b_E\cdot p_E)\nonumber \\
&&+48k_{E}^{\mu}p_{E}^{\nu}(b_E\cdot p_E)(b_E\cdot k_E)+48p_{E}^{\mu}k_{E}^{\nu}(b_E\cdot p_E)(b_E\cdot k_E)+48(b_E\cdot k_E)^{2}p_{E}^{2}\delta^{\mu\nu}\nonumber \\
&&-16m^{2}b_{E}^{\mu}b_{E}^{\nu}p_{E}^{2}+48m^{2}b_{E}^{2}p_{E}^{2}\delta^{\mu\nu}+24k_{E}^{\mu}k_{E}^{\nu}b_{E}^{2}p_{E}^{2}+48k_{E}^{\mu}b_{E}^{\nu}(b_E\cdot k_E)p_{E}^{2}\nonumber \\
&&+48b_{E}^{\mu}k_{E}^{\nu}(b_E\cdot k_E)p_{E}^{2}+24k_{E}^{\mu}p_{E}^{\nu}(b_E\cdot k_E)^{2}+24p_{E}^{\mu}k_{E}^{\nu}b_{E}^{2}(p_E\cdot k_E)\nonumber \\
&&+64k_{E}^{\mu}b_{E}^{\nu}(b_E\cdot p_E)(p_E\cdot k_E)+64b_{E}^{\mu}k_{E}^{\nu}(b_E\cdot p_E)(p_E\cdot k_E)+48p_{E}^{\mu}b_{E}^{\nu}(b_E\cdot k_E)(p_E\cdot k_E)\nonumber \\
&&+48b_{E}^{\mu}p_{E}^{\nu}(b_E\cdot k_E)(p_E\cdot k_E)+96(b_E\cdot p_E)(b_E\cdot k_E)(p_E\cdot k_E)\delta^{\mu\nu}],
\end{eqnarray}
\begin{eqnarray}
\Pi_3^{\mu\nu} &=&-\int\frac{d^4k_E}{(2\pi)^4}\frac{1}{(k_E^2+m^2)^5}[192(b_E\cdot k_E)^{2}(p_E\cdot k_E)^{2}\delta^{\mu\nu}+192m^{2}b_{E}^{2}(p_E\cdot k_E)^{2}\delta^{\mu\nu}\nonumber \\
&&+128k_{E}^{\mu}k_{E}^{\nu}b_{E}^{2}(p_E\cdot k_E)^{2}+256k_{E}^{\mu}b_{E}^{\nu}(b_E\cdot k_E)(p_E\cdot k_E)^{2}\nonumber \\
&&+256b_{E}^{\mu}k_{E}^{\nu}(b_E\cdot k_E)(p_E\cdot k_E)^{2}+192k_{E}^{\mu}k_{E}^{\nu}(b_E\cdot k_E)^{2}p_{E}^{2}+192m^{2}k_{E}^{\mu}k_{E}^{\nu}b_{E}^{2}p_{E}^{2}\nonumber \\
&&+192k_{E}^{\mu}p_{E}^{\nu}(b_E\cdot k_E)^{2}(p_E\cdot k_E)+192p_{E}^{\mu}k_{E}^{\nu}(b_E\cdot k_E)^{2}(p_E\cdot k_E)\nonumber \\
&&+192m^{2}k_{E}^{\mu}p_{E}^{\nu}b_{E}^{2}(p_E\cdot k_E)+192m^{2}p_{E}^{\mu}k_{E}^{\nu}b_{E}^{2}(p_E\cdot k_E)\nonumber \\
&&+512k_{E}^{\mu}k_{E}^{\nu}(b_E\cdot p_E)(b_E\cdot k_E)(p_E\cdot k_E)],
\end{eqnarray}
\begin{eqnarray}
\Pi_4^{\mu\nu} &=&\int\frac{d^4k_E}{(2\pi)^4}\frac{1280}{(k_E^2+m^2)^6}[k_{E}^{\mu}k_{E}^{\nu}(b_E\cdot k_E)^{2}(p_E\cdot k_E)^{2}+m^{2}k_{E}^{\mu}k_{E}^{\nu}b_{E}^{2}(p_E\cdot k_E)^{2}].
\end{eqnarray}

In order to effectively tackle the issue of the finite temperature behavior of the above expressions, we separate the space and time components of the four-momentum $k^\sigma_E=(k_0,\vec{k})$ as $k^\sigma_E\rightarrow \hat{k}^\sigma +k_0\delta^{\sigma 0}$, so that $ \hat{k}^\sigma=(0,\vec{k})$. Also, due to the symmetry of the integral under spacial rotations, it is possible to use the substitutions
\begin{eqnarray}
\hat{k}^\alpha \hat{k}^\beta \rightarrow \frac{\hat{k}^2}{d}(\delta^{\alpha\beta}-\delta^{\alpha 0} \delta^{\beta 0})
\end{eqnarray}
and
\begin{eqnarray}
\hat{k}^\alpha \hat{k}^\beta \hat{k}^\delta \hat{k}^\gamma &\rightarrow& \frac{\hat{k}^4}{d(d+2)}[(\delta^{\alpha\beta}-\delta^{\alpha 0}\delta^{\beta 0})(\delta^{\delta\gamma}-\delta^{\delta 0}\delta^{\gamma 0})+(\delta^{\alpha\delta}-\delta^{\alpha 0}\delta^{\delta 0})(\delta^{\beta\gamma}-\delta^{\beta 0}\delta^{\gamma 0})\nonumber\\ 
&&+(\delta^{\alpha\gamma}-\delta^{\alpha 0}\delta^{\gamma 0})(\delta^{\beta \delta}-\delta^{\beta 0}\delta^{\delta 0})].
\end{eqnarray}
These procedures lead to the result
\begin{eqnarray}\label{sumprojs}
\Pi^{\mu\nu}&=&A ((b_E\cdot p_E)^2 \delta^{\mu\nu}+p_E^2 b_E^\mu b_E^\nu-(b_E\cdot p_E)b_E^\mu p_E^\nu-(b_E\cdot p_E)p_E^\mu b_E^\nu-b_E^2p_E^2 \delta^{\mu\nu}+b_E^2 p_E^\mu p_E^\nu) \nonumber\\
&&+B(b_0^2p_0^2\delta^{\mu\nu}+b_0^2p_E^2\delta^{\mu0}\delta^{\nu0}-b_0^2p_0 p_E^\mu\delta^{\nu0}-b_0^2p_0 \delta^{\mu0}p_E^\nu) \nonumber\\
&&+C(b_E^2p_0^2\delta^{\mu\nu}+b_E^2p_E^2\delta^{\mu0}\delta^{\nu0}-b_E^2p_0\delta^{\mu0}p_E^\nu-b_E^2p_0p_E^\mu\delta^{\nu0}) \nonumber\\
&&+D(b_0^2p_E^2\delta^{\mu\nu}-b_0^2p_E^\mu p_E^\nu),
\end{eqnarray}
where
\begin{eqnarray}
A&=&-\frac{1}{3}2^{3-d}m^{2}\pi^{-\frac{d}{2}}\Gamma\left(4-\frac{d}{2}\right)\int\frac{dk_0}{2\pi}(k_0^{2}+m^{2})^{\frac{d}{2}-4}\\
B&=&-\frac{1}{3}2^{1-d}\pi ^{-\frac{d}{2}}\Gamma\left(3-\frac{d}{2}\right)\int\frac{dk_0}{2\pi}(k_0^{2}+m^{2})^{\frac{d}{2}-5} \nonumber\\
&&\times(3m^{4}+(d-5)k_0^{2}(6m^{2}+(d-3)k_0^{2}))\\
C&=&\frac{1}{3}2^{2-d}m^{2}\pi^{-\frac{d}{2}}\Gamma\left(4-\frac{d}{2}\right)\int\frac{dk_0}{2\pi}(m^{2}+k_0^{2})^{\frac{d}{2}-5}(m^{2}+(d-7)k_0^{2})\\
D&=&-\frac{1}{3}2^{2-d}\pi^{-\frac{d}{2}}\Gamma\left(3-\frac{d}{2}\right)\int\frac{dk_0}{2\pi}(m^{2}+k_0^{2})^{\frac{d}{2}-4}(m^{2}+(d-5)k_0^{2}).
\end{eqnarray}
One can straightforwardly verify that each of four contributions to the expression (\ref{sumprojs}), proportional to $A$, $B$, $C$, and $D$ respectively, is separately gauge invariant. Therefore, the gauge invariance will be automatically maintained at the finite temperature as well.
If we then calculate the $k_0$ integrals in these expressons, we get $A=-\frac{1}{6m^2\pi^2}$ and $B=0=C=D$, as expected. 

We are now in position to implement the finite temperature through employing the Matsubara formalism, which consist in taking $k_0=(n+1/2)2\pi/\beta$ and changing $(1/2\pi)\int dk_0\to 1/\beta \sum_n$. By performing the summations, we obtain
\begin{eqnarray}
\label{termcoef}
A&=&-\frac{1}{6m^2\pi^{2}}-\frac{1}{6m^2}\int_{|\xi|}^{\infty}dz\frac{(\xi^{2}-2z^{2})}{(z^2-\xi^2)^{1/2}}\sech^{2}(\pi z)\tanh(\pi z), \\
B&=& -C = \frac{1}{12m^2}\pi^{2}\xi^{2}\int_{|\xi|}^{\infty}dz(z^2-\xi^2)^{1/2}\,\sech^{5}(\pi z)(\sinh(3\pi z)-11\sinh(\pi z)), \\
D&=&\frac{1}{6m^2}\xi^{2}\int_{|\xi|}^{\infty}dz(z^2-\xi^2)^{-1/2}\,\sech^{2}(\pi z)\tanh(\pi z).
\end{eqnarray}
We observe that in the high temperature limit, $\xi\to0$, all the above coefficients vanish.

The contribution to the effective action, corresponding to (\ref{sumprojs}) with the temperature dependent coefficients $A,B,C$, and $D$, looks like
\bea
\label{finmin}
S_{AA}=A(2b^2F_{\mu\nu}F^{\mu\nu}-4b^{\mu}F_{\mu\nu}b_{\lambda}F^{\lambda\nu})+Bb_i^2F_{0\mu}F^{0\mu}
+Db^2_0F_{\mu\nu}F^{\mu\nu}.
\eea
We see that this result involves, first, the usual aether-like term, second, the Lorentz-invariant Maxwell term accompanied by constant multiplier (which involves both $b^2$ and $b^2_0$), third, the term $F_{0\mu}F^{0\mu}$ which can be treated as a specific form of the aether term corresponding to the case when the vector $b^{\mu}$ is purely space-like.

\subsection{Minimal-nonminimal contribution}

Now, let us consider the ``mixed'' Feynman diagrams, involving both minimal and nonminimal couplings, and depicted at Fig.1.

\begin{figure}[ht] 
\includegraphics[scale=0.6]{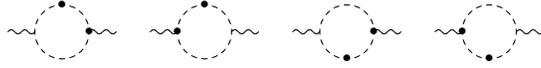}
\caption{\small{Contributions to the two-point function of the vector field.}}
\label{figure9}
\end{figure}

Similarly to the calculations in \cite{MNP}, here we consider the Lorentz-breaking insertions $\gamma_5\bs$ introduced both in the vertices and in the propagators (both these insertions are denoted by the symbol $\bullet$).
The calculations do not essentially differ from those ones carried out in \cite{MNP}, since the loop integrals and traces are identically the same (actually, the only difference from those papers is related to the fact that in one of the vertices, the $A_{\mu}$ field is replaced by its ``dual'' $\epsilon_{\mu\nu\lambda\rho}b^{\nu}F^{\lambda\rho}$). As a result, we arrive at
\begin{eqnarray}
\label{Ef3}
S_{AF}&=&eg\int d^4x\;
I_{\rho}\epsilon^{\rho\nu\lambda\mu}
(\epsilon_{\nu\alpha\beta\gamma}F^{\alpha\beta}b^{\gamma} \partial_{\lambda}A_{\mu}+ 
A_{\nu}\epsilon_{\mu\kappa\eta\sigma}\partial_{\lambda}F^{\kappa\eta}b^\sigma),
\end{eqnarray}
where $k_{\rho}$ is a constant vector whose explicit form is
\begin{eqnarray}
\label{I1}
I_{\rho}=2i\int\,\frac{d^4k}{(2\pi)^4}\,\frac{b_{\rho}(k^2+3m^2)-4k_{\rho}(b\cdot k)}{(k^2-m^2)^3},
\end{eqnarray}
which can be calculated through different regularization schemes (a very incomplete list of the approaches to calculating this vector which is equivalent to the calculation of the Carroll-Field-Jackiw term can be found in \cite{list}).

So, let us discuss this result at the finite temperature, at least for two methods of calculations, having a mere purpose to illustrate the ambiguity of the results for this vector. Within the simplest approach, we can follow the line elaborated in \cite{ours} and repeat all calculations carried out there. In \cite{ours}, two schemes were used to obtain the $I_{\rho}$. Within the first scheme, one deals with the expression (\ref{I1}) and considers separately the time-like and the space-like parts of the $b_{\mu}$, which after corresponding symmetrizations allows to obtain
\bea
\label{sche1}
I_0&=&\frac{1}{4}b_0;\nonumber\\
I_i&=&\frac{1}{4}b_i\left[\frac{1}{\pi^2}+\frac{1}{2}F_5(\xi)\right],
\eea
where $\xi=\frac{m}{2\pi T}$, and the function $F_5(\xi)$ has been discussed in details in \cite{ours} and yields
\bea
F_5(\xi)=\int\limits_{|\xi|}^{\infty}dz(z^2-\xi^2)^{1/2}\,\sech^2(\pi z)\tanh (\pi z).
\eea
Within another scheme, one first imposes the spherical symmetry by making the replacement $k_{\rho}k^{\nu}\to \frac{1}{4}\delta^{\nu}_{\rho} k^2$ in (\ref{I1}) which allows to rewrite (\ref{I1}) as
\begin{eqnarray}
\label{I1a}
I_{\rho}=6ib_{\rho}\int\,\frac{d^4k}{(2\pi)^4}\,\frac{m^2}{(k^2-m^2)^3},
\end{eqnarray}
which yields
\bea
\label{sche2}
I_0&=&\frac{3}{16}b_0\left[\frac{1}{2\pi^2}+F_5(\xi)\right];\nonumber\\
I_i&=&\frac{3}{16}b_i\left[\frac{1}{2\pi^2}+F_5(\xi)\right].
\eea
So, within this approach the vector $I_{\mu}$ is proportional to $b_{\mu}$ just as in the zero temperature case.
More details of these calculations can be found in \cite{ours}. 
In principle, we can, alternatively, follow the approach developed in \cite{ours1} based on the formalism proposed in \cite{Seminara}, and obtain more possibilities for these vectors (we note this list of calculation schemes is not an exhaustive one, and, in principle, other values for the vector $I^{\mu}$ can be found as well). Thus, we demonstrated that this source for the ambiguity of the aether term still works at the finite temperature.

It remains to substitute these values for $I_{\mu}$ into the expression (\ref{Ef3}) for the aether term. After the contraction of two Levi-Civita symbols, one has
\begin{eqnarray}
\label{finmixed}
S_{AF}&=&eg\int d^4x\;
\Big[2F_{\mu\nu}F^{\mu\nu}(b\cdot I)-4F^{\alpha\beta}F_{\alpha\gamma}I_{\beta}b^{\gamma}
\Big],
\end{eqnarray}
where the temperature dependent vector $I_{\beta}$ is ambiguous, in some schemes of calculation it turns out to be proportional to the vector $b^{\mu}$ and can be read off from the expressions (\ref{sche1}) or (\ref{sche2}), depending on the scheme of calculations we choose. The first, Lorentz-invariant term is also ambiguous. Therefore we demonstrated the finiteness and ambiguity for the contribution to the aether term from this sector. 

To close the discussion, we note that the complete result for the aether-like term in the finite temperature case is represented by a sum of one of the contributions in (\ref{finnonmin}) with the expression (\ref{finmin}) and the result (\ref{finmixed}) in which either the form (\ref{sche1}) or the form (\ref{sche2}) of the vector $I_{\mu}$ is used. Therefore, we conclude that the aether-like term in the finite temperature case involves two ambiguities as well as in the zero temperature case. 

\section{Summary}

We have showed that the aether-like term, known to be finite and ambiguous at the zero temperature, is finite and ambiguous at the finite temperature as well, with two ambiguities occur, one in the purely nonminimal sector, where we have at least four consistent schemes for calculation of the aether term, and another in the mixed sector where the ambiguity is just the same as in the case of the CFJ term, and it is not eliminated by the finite temperature, while within this paper we have demonstrated only two possible schemes for its calculations based on an appropriate adaptation of results of the paper \cite{ours} which clearly yield different results. In principle, our list of prescriptions for calculations of the aether term in the Lorentz-breaking QED with a magnetic-like nonminimal coupling is not exhaustive. We notice once more that, in the purely nonminimal sector, the aether-like contributions are superficially divergent, but in any of these prescriptions they turn out to be finite where the zero-temperature divergences in all cases are cancelled due to the properties of the Euler gamma function.

Since, as it is claimed in \cite{JackAmb}, the ambiguities of the results for the CFJ term are related with the ABJ anomaly, we close our paper with the statement that the ABJ anomaly (which as we already noted \cite{MNP} apparently can be extended by introducing the higher-derivative terms), and, possibly, one more anomaly responsible for the ambiguity (and finiteness) of the aether term in a purely nonminimal sector, stay as well in the zero temperature case. However, this probably existing anomaly still requires a detailed study.

{\bf Acknowledgements.} This work was partially supported by Conselho
Nacional de Desenvolvimento Cient\'{\i}fico e Tecnol\'{o}gico (CNPq). The work by A. Yu. P. has been supported by the
CNPq project No. 303438/2012-6.


\begin{thebibliography}{20}
\bibitem{JackAmb} R. Jackiw, Int. J. Mod. Phys. B {\bf 14}, 2011 (2000), hep-th/9903044.
\bibitem{CFJ} S. Carroll, G. Field, R. Jackiw, Phys. Rev. D {\bf 41}, 1231 (1990).
\bibitem{Chung} J. M. Chung, Phys. Rev. D {\bf 60}, 127901 (1999), hep-th/9904037.
\bibitem{Seminara} L. Cervi, L. Griguolo, D. Seminara, Phys. Rev. D {\bf 64}, 105003 (2001), hep-th/0104022. 
\bibitem{ours} M.~Gomes, J.~R.~Nascimento, E.~Passos, A.~Yu.~Petrov and A.~J.~da Silva, Phys. Rev. D {\bf 76}, 047701 (2007), arXiv: 0704.1104. 
\bibitem{ours1} J.~R.~Nascimento, E.~Passos, A.~Yu.~Petrov and F.~A.~Brito, JHEP {\bf 0706}, 016 (2007), arXiv: 0705.1338.
\bibitem{aether} M. Gomes, J. R. Nascimento, A. Yu. Petrov, A. J. da Silva, Phys. Rev. D {\bf 81}, 045018 (2010), arXiv: 0911.3548; ``On the Aether-like Lorentz-breaking action for the electromagnetic field'', arXiv: 1008.0607. 
\bibitem{MNP} T. Mariz, J. R. Nascimento, A. Yu. Petrov, Phys. Rev. D {\bf 85}, 125003 (2012), arXiv: 1111.0198.
\bibitem{MP} R. Myers, M. Pospelov, Phys. Rev. Lett. {\bf 90}, 211601 (2003), hep-ph/0301124.
\bibitem{LMS} J. Leite, T. Mariz, Europhys. Lett. {\bf 99}, 21003 (2012), arXiv: 1110.2127; J. Leite, T. Mariz, W. Serafim, J. Phys. G {\bf 40}, 075003 (2013).
\bibitem{Reyes} C. M. Reyes, L. F. Urrutia, J. D. Vergara, Phys.Rev. D78, 125011 (2008), arXiv: 0810.5379; C. M. Reyes, Phys. Rev. D82, 125036 (2010), arXiv: 1011.2971; Phys. Rev. D87, 125028 (2013), arXiv: 1307.5340.
\bibitem{Kostel} V. A. Kostelecky, Phys. Rev. D69, 105009 (2004), hep-th/0312310.
\bibitem{Carroll} S. Carroll, H. Tam, Phys. Rev. D {\bf 78}, 044047 (2008), arXiv: 0802.2521.
\bibitem{Almeida} V. Santos, C. A. S. Almeida, Phys. Lett. B718, 1114 (2013), arXiv: 1211.4542.
\bibitem{aethercl} H. Belich, T. Costa-Soares, M. M. Ferreira Jr., J. A. Helayel-Neto, Eur. Phys. J. C41, 421 (2005), hep-th/0410104; Eur. Phys. J. C42, 127 (2005), hep-th/0411151; H. Belich, T. Costa-Soares, M. M. Ferreira Jr., J. A. Helayel-Neto, F. M. O. Moucherek, Phys. Rev. D74, 065009 (2006), hep-th/0604149; H. Belich, L. P. Colatto, T. Costa-Soares, J. A. Helayel-Neto, M. T. D. Orlando, Eur. Phys. J. C62, 425 (2009), arXiv: 0806.1253; R. Casana, M. M. Ferreira, C. Santos, Phys. Rev. D78, 105014 (2009), arXiv: 0810.2817; R. Casana, M. M. Ferreira, J. S. Rodrigues, M. R. O. Silva, Phys. Rev. D80, 085026 (2009), arXiv: 0907.1924;  R. Casana, A. R. Gomes, M. M. Ferreira, P. R. D. Pinheiro, 
Phys. Rev. D80, 125040 (2009), arXiv: 0909.0544; 
F. R. Klinkhammer, M. Schreck, Nucl. Phys. B848, 90 (2011), arXiv: 1011.4258; Nucl. Phys. B856, 666 (2012), arXiv: 1110.4101; M. Schreck, Phys. Rev. D86, 065038 (2011), arXiv: 1111.4182;
R. Casana, E. S. Carvalho, M. M. Ferreira, 
Phys. Rev. D84, 045008 (2011), arXiv: 1107.2664; R. Casana, M. M. Ferreira, R. P. M. Moreira, 
Phys. Rev. D84, 125014 (2011), arXiv: 1108.6193.
\bibitem{Clovis} M. S. Guimaraes, J. R. Nascimento, A. Yu. Petrov, C. Wotzasek, Europhys. Lett. 95, 51002 (2011), arXiv: 1010.3666.
\bibitem{aether1} G. Gazzola, H. G. Fargnoli, A. P. Baeta Scarpelli, M. Sampaio, M. C. Nemes, J. Phys. G {\bf 39}, 035002 (2012), arXiv: 1012.3291;
A. P. Baeta Scarpelli, T. Mariz, J. R. Nascimento, A. Yu. Petrov, Eur. Phys. J. C{\bf 73}, 2526 (2013), arXiv: 1304.2256.
\bibitem{For}L.~H.~Ford, Phys. Rev. D {\bf 21}, 933 (1980).
\bibitem{list} R. Jackiw, V. A. Kostelecky, Phys. Rev. Lett. 82, 3572 (1999) [hep-ph/9901358];
S. Coleman and S. L. Glashow, Phys. Rev. D {\bf59}, 116008 (1999) [hep-ph/9812418];
M. P\'erez-Victoria, Phys. Rev. Lett. {\bf83}, 2518 (1999) [hep-th/9905061];
J. M. Chung and P. Oh, Phys. Rev. D {\bf60}, 067702 (1999) [hep-th/9812132];
J. M. Chung, Phys. Rev. D {\bf60}, 127901 (1999) [hep-th/9904037];
W. F. Chen, Phys. Rev. D {\bf60}, 085007 (1999) [hep-th/9903258];
J. M. Chung, Phys. Lett. B {\bf461}, 138 (1999) [hep-th/9905095];
C.~Adam and F.~R.~Klinkhamer, Phys.\ Lett.\ B {\bf 513}, 245 (2001) [hep-th/0105037];
G. Bonneau, Nucl. Phys. B {\bf593}, 398 (2001) [hep-th/0008210];
Yu. A. Sitenko, Phys. Lett. B {\bf515}, 414 (2001) [hep-th/0103215];
M. Chaichian, W. F. Chen, and R. Gonz\'alez Felipe, Phys. Lett. B {\bf 503}, 215 (2001) [hep-th/0010129];
J. M. Chung and B. K. Chung, 105015 (2001) [hep-th/0101097]; Yu.~A.~Sitenko and K.~Y.~Rulik, Eur.\ Phys.\ J.\  C {\bf 28}, 405 (2003) [hep-th/0212007];
D.~Bazeia, T.~Mariz, J.~R.~Nascimento, E.~Passos and R.~F.~Ribeiro, J.\ Phys.\ A {\bf 36}, 4937 (2003) [hep-th/0303122];
Y.~L.~Ma and Y.~L.~Wu, Phys.\ Lett.\  B {\bf647}, 427 (2007) [hep-ph/0611199];
A. A. Andrianov, P. Giacconi, and R. Soldati, JHEP {\bf 0202}, 030 (2002) [hep-th/0110279].
\end{thebibliography}
\end{document}